%% file: main.tex
\documentclass[acmsmall,screen]{acmart}

\AtBeginDocument{%
  }

\setcopyright{acmlicensed}
\copyrightyear{2018}
\acmYear{2018}
\acmDOI{XXXXXXX.XXXXXXX}

\acmConference[Conference acronym 'XX]{Make sure to enter the correct
  conference title from your rights confirmation emai}{June 03--05,
  2018}{Woodstock, NY}
\acmISBN{978-1-4503-XXXX-X/18/06}

\usepackage{enumitem}
\usepackage{bm}
\usepackage{caption}
\usepackage{natbib}
\usepackage{comment}
\usepackage{soul}
\usepackage{enumerate}
\usepackage{adjustbox}
\usepackage{subcaption}
\usepackage{relsize}
\usepackage{xcolor}
\usepackage{multirow} 
\usepackage[linesnumbered,ruled,vlined]{algorithm2e}
\usepackage{amsmath}
\usepackage{graphicx}
\usepackage{multirow}
\usepackage{booktabs}

\input{macro.tex}
\begin{document}

\title{\ourtool{}:Consistency-Augmented Iterative Interaction
Framework to Enhance the Reliability of Code Generation}

\author{Jinhao Dong}
\authornote{This work was done when Jinhao Dong was a visiting student in National University of Singapore (NUS) and Singapore Management University (SMU).}
\affiliation{%
  \institution{Peking University}
  \city{Beijing}
  \country{China}
}
\email{dongjinhao@stu.pku.edu.cn}

\author{Jun Sun}

\affiliation{%
  \institution{Singapore Management University}
  \country{Singapore}
}
\email{junsun@smu.edu.sg}

\author{Wenjie Zhang}

\affiliation{%
  \institution{National University of Singapore}
  \country{Singapore}
}
\email{wjzhang@nus.edu.sg}

\author{Jin Song Dong}
\affiliation{%
  \institution{National University of Singapore}
  \country{Singapore}
}
\email{dcsdjs@nus.edu.sg}
\author{Dan Hao}
\affiliation{%
  \institution{Peking University}
  \city{Shenzhen}
  \country{China}
}
\email{haodan@pku.edu.cn}
\begin{abstract}
Code generation techniques generate code snippets automatically based on the problem requirements in natural language, which has the potential to significantly improve the developer's productivity. Recently, large language models (\llm{}s) achieve the \sota{} performance on code generation, which are pre-trained on extensive code-specific corpora. 
However, LLMs still struggle at times to generate accurate code, which diminishes their promised efficiency as developers must spend significant effort evaluating and debugging the generated code. To improve the reliability and quality of the generated codes, researchers propose to leverage \self{} to obtain a better code based on generating and ranking multiple candidates. The existing approach is problematic as  \self{} thinks a code is better when (1) the code pass more tests (inter-consistency) (2) more codes share the same behavior (intra-consistency). However, because the tests are also generated by \llm{}s, they could be wrong as well. As a result, majority voting based on testing results is unreliable. 
Relying solely on consistency is insufficient to address this issue; integrating user feedback is essential for effectively guiding consistency. We show that with minimal human effort, performance can be significantly enhanced.

We propose Consistency-Augmented Iterative Interaction
Framework to Enhance the Reliability of Code Generation, \ourtool{}, which is an approach that aims to improve the performance of a code generator through two distinctive ingredients, i.e., (1) lightweight user effort for validating the correctness of selected tests; and (2) a dynamic strategy for ranking, localizing and correcting multiple tests and codes.
Overall, we propose a lightweight interaction framework that incorporates user feedback to correct identified tests and guide the iterative process. The iteration rounds are only $4$ in average with the help of consistency. With only lightweight human efforts, we can achieve an improvement of 33\% towards the base model. 
In each iteration, we propose a \threestage{} co-evolution process between codes and tests. The co-evolution process improves the quality of codes and tests iteratively, which makes both the consistency voting from codes to tests and the consistency voting from tests to codes more reliable. 
We conduct a comprehensive evaluation on \ourtool{}. Firstly, we evaluate the effectiveness of \ourtool{} on improving the \llm{} code generation. We conduct two simulated experiments to automatically evaluate \ourtool{} where (1) \oo{} is used to simulate user feedback and (2) \gtsolution{} is used to simulate user feedback, enabling extensive quantitative analysis.  Using a suboptimal model, \gptthree{}, \ourtool{} achieves an average improvement of 32.9\% over \gptthree{}, a 11.1\% improvement over the \sotasimple{} post-processing technique, \mpsc{}, and a 12.32\% improvement compared to the most advanced general \llm{}, \gptfour{}. This improvement is achieved with only a 4-round interaction with users, requiring minimal user effort. Furthermore, \ourtool{} also achieves consistent improvements when built on the \sotasimple{} \llm{} \gptfour{} and even on the reasoning \llm{} \oo{}. Moreover, we conduct a user study and we also explore the overhead of \ourtool{} from time and cost.


\end{abstract}

\ccsdesc[500]{Software and its engineering~Automatic programming}
\ccsdesc[500]{Software and its engineering~Software development techniques}

\keywords{Code Generation, Self-Consistency, Iterative Interaction}

\maketitle

\section{Introduction}
\label{sec:intro}
\input{sections/introduction}
\section{Motivation}
\label{sec:motivation}
\input{sections/motivation}

\section{Approach}
\label{sec:approach}
\input{sections/approach}

\section{Experimental Setup}
\label{sec:setup}
\input{sections/setup}

\section{Results and Analysis}
\label{sec:results}
\input{sections/results}

\section{Discussion}
\label{sec:discussion}
\input{sections/discussion}

\section{Related Work}
\label{sec:related}
\input{sections/relatedwork}

\section{Conclusion}
\label{sec:conclusion}
\input{sections/conclusion}


\bibliographystyle{ACM-Reference-Format}
\bibliography{ref}


\end{document}

%% file: macro.tex
\newcommand{\sota}{state-of-the-art}
\newcommand{\sotasimple}{SOTA}
\newcommand{\self}{\textit{Consistency}}
\newcommand{\diversity}{\textit{diversity}}
\newcommand{\llm}{LLM}
\newcommand{\codet}{C\textsc{ode}T}
\newcommand{\mpsc}{MPSC}
\newcommand{\spec}{specification}
\newcommand{\ourtool}{C\textsc{on}AIR}
\newcommand{\ourtoolgt}{$\text{\ourtool}_{\text{GT}}$}
\newcommand{\ourtooloo}{$\text{\ourtool}_{\text{o1}}$}
\newcommand{\openai}{OpenAI}
\newcommand{\oo}{OpenAI o1}
\newcommand{\gptthree}{GPT-3.5}
\newcommand{\gptfour}{GPT-4o}
\newcommand{\humaneval}{HumanEval}
\newcommand{\mbpp}{MBPP}
\newcommand{\humanevalplus}{HumanEval+}
\newcommand{\pre}{prerequisite}
\newcommand{\code}{$c$}
\newcommand{\fixcode}{$c_\text{fix}$}
\newcommand{\codetext}{c}
\newcommand{\bestcode}{$\hat{c}$}
\newcommand{\des}{$d$}
\newcommand{\test}{$t$}
\newcommand{\testtext}{t}
\newcommand{\codeset}{$\mathbf{C}$}
\newcommand{\discodeset}{$\mathbf{C}_\text{dis}$}

\newcommand{\remcodeset}{$\mathbf{C}_\text{rem}$}

\newcommand{\codesettext}{\mathbf{C}}
\newcommand{\testset}{$\mathbf{T}$}

\newcommand{\model}{$\mathcal{M}$}
\newcommand{\feedback}{$f$}
\newcommand{\feedbackset}{$\mathbf{F}$}
\newcommand{\chatgpt}{ChatGPT}
\newcommand{\concodetest}{$\texttt{Con}_{c\rightarrow{}t}$}
\newcommand{\concodetesttext}{\texttt{Con}_{c\rightarrow{}t}}
\newcommand{\contestcode}{$\texttt{Con}_{t\rightarrow{}c}$}

\newcommand{\context}{\texttt{Con}}
\newcommand{\threestage}{rank-correct-fix}
\newcommand{\cortestset}{$\mathbf{T}_\text{cor}$}
\newcommand{\cortest}{$t_\text{cor}$}

\newcommand{\unktestset}{$\mathbf{T}_\text{unk}$}

\newcommand{\wrongtest}{$t_\text{w}$}

\newcommand{\passk}{Pass@k}
\newcommand{\passone}{Pass@1}
\newcommand{\passtwo}{Pass@2}
\newcommand{\passfive}{Pass@5}

\newcommand{\gtsolution}{ground truth solution}
\newcommand{\indicator}{consistency indicator}
\newcommand{\functionequal}{functionally equivalent}
\newcommand{\samecode}{counterpart}

\DeclareMathOperator*{\argmax}{argmax}

\SetKwComment{Comment}{$\triangleright$\ }{}
\SetNlSty{textnormal}{}{}

%% file: sections/introduction.tex
Code generation techniques automatically generate code snippets that implement desired functionality based on natural language requirements. These techniques can reduce the effort required by developers to write code and improve development productivity, as extensively studied in the literature~\cite{sun2020treegen,kang2023large,li2023codeeditor}. Recent progress in large language models (LLMs) have significantly impacted the field of code generation. Researchers have introduced various LLMs~\cite{luo2023wizardcoder,achiam2023gpt,guo2024deepseek,nijkamp2022codegen,du2021glm,li2023starcoder,fried2022incoder, hui2024qwen2} (e.g., GPT-4~\cite{achiam2023gpt}, DeepSeek-Coder~\cite{guo2024deepseek}, and CodeGen~\cite{nijkamp2022codegen}) that achieve \sotasimple{} performance, due to their massive parameter scales and pre-training on extensive code-specific corpora.

Although LLMs have demonstrated impressive performance, their outputs are not always reliable. Enhancing the reliability of LLM-generated results is crucial. This is because it often takes significant effort for developers to understand and correct the generated code if it turns out to be wrong. To increase the reliability of LLM-generated results, \self{}~\cite{wang2022self,sun2022recitation,chen2022codet,huang2023enhancing,xiong2023examining,zhang2023algo} is proposed as an effective and lightweight technique that generates multiple solutions in various ways for each input query, then determines the final answer through a majority vote to ensure the most consistent result. \self{} is founded on the assumption that the tasks generally have multiple reasoning paths leading to a correct answer~\cite{stanovich2000advancing}. \self{} is based on the concept of \diversity{}. When diverse approaches lead to the same answer, that consistent result is likely correct, as the chance of multiple perspectives producing the same error is low. \self{} helps mitigate the randomness, thereby enhancing output reliability and significantly boosting performance~\cite{wang2022self,wang2024soft,sun2022recitation,xiong2023examining}. Other post-processing techniques that aim to enhance the reliability of LLMs are generally more resource-intensive compared to \self{}. For instance, some methods involve training an additional verifier~\cite{cobbe2021training,ni2023lever} to validate the outputs of LLMs or an extra re-ranker~\cite{thoppilan2022lamda} to prioritize results. These approaches require the training of a separate model, and the reliability of its verification results may still be questionable. In contrast, \self{} operates without necessitating further training or auxiliary models. Recently, researchers have applied \self{} to code generation tasks~\cite{chen2022codet,huang2023enhancing}. Coding tasks often involve diverse approaches, including various APIs, algorithms, data structures, and programming paradigms (e.g., procedural, object-oriented, and functional programming), which provide a range of perspectives for applying consistency.

While \self{} improves the reliability of the codes generated by LLMs, the existing approaches share one common limitation. \textit{Existing techniques overlook the preconditions of using \self{}, and \self{} alone is insufficient to guarantee the reliability of \llm{}s.} In particular, the precondition of using \self{} is that the \indicator{} (which is used to assess consistency) has relatively good quality. Only in this way, the results of majority voting based on \indicator{}s is trustworthy and the consistent behavior is indeed correct. Because the \indicator{}s are also generated by LLMs, they could be wrong as well. As a result, \self{} and majority voting based on incorrect \indicator{}s are unreliable. Relying solely on consistency is insufficient to address this issue. Engaging with users and utilizing their feedback is necessary to guide consistency effectively. We demonstrate that with minimal human efforts, performance can be significantly improved.
Specifically, in code generation, current consistency-based LLM approaches typically utilize tests or specifications as \indicator{}s~\cite{chen2022codet,huang2023enhancing}. These methods identify code that passes the most tests (termed inter-consistency) and has the highest number of functionally equivalent counterparts, as indicated by passed tests (termed intra-consistency). As shown in Fig.~\ref{fig:moti_exec} (which will be discussed in detail in Section~\ref{sec:motivation}), codes in group 1 are selected because they pass more tests and have more functionally equivalent counterparts that pass the same set of tests. Tests serve as \indicator{}, assessing the consistency level of the generated code and supporting a majority voting process. However, because the tests are also generated by
LLMs, they could also be wrong. The consistency derived from these buggy tests and majority voting based on testing results is unreliable.
A buggy code that passes a higher number of erroneous tests may be mistakenly considered as reliable or correct. In our experiments, we compute that the tests generated by \llm{}s have an average error rate of 37.7\% across three widely used code generation datasets: \humaneval{}, \humanevalplus{}, and \mbpp{}. The high error rate of generated tests poses a significant threat to the quality of \self{}. However, existing techniques neglect this issue when implementing \self{}. It is therefore essential to enhance the quality of \indicator{} prior to leveraging consistency. 
\textit{Furthermore, existing techniques make limited use of \self{}.} Relying solely on majority voting provides only a superficial application of \self{}. There are additional ways to leverage \self{}, such as using inconsistency to identify potential problems. Once identified, these problems can be fixed to enhance reliability. Moreover, all existing consistency-based techniques lack post-processing steps and simply select a single output. However, the candidate codes might still need additional adjustments.

In this work, we propose \ourtool{}, i.e., Consistency-Augmented Iterative Interaction
Framework to Enhance the Reliability of Code Generation, which incorporates a lightweight interaction framework to gather user feedback and a co-evolution process to iteratively enhance the quality of both tests and codes. \ourtool{} has two ingredients that distinguish it from existing approaches. \textit{Firstly, the developers are involved as the ultimate oracle for two reasons.} The first reason is that, in many cases, only the developer knows whether certain testing result is correct or not. Since both the tests and code are generated by LLMs, they may each contain issues, making it impossible to determine correct test outputs based on them alone. In real-world development, developers must write tests to ensure the correctness of the codes. The second reason is that, keeping the developer in the loop nurtures certain code ownership and makes sure that the developer always has a good view of the code generation process. In coding, the process often holds more significance than the result. This involvement allows developers to gain a deeper knowledge of the code’s details, enabling them to respond more swiftly to future bugs. However, it is important to consult the developer minimally, e.g., only requiring them to check the validity of the testing results rather than the code itself, and limiting checks to the fewest possible rounds. Following these principles, we propose a lightweight interaction framework that incorporates user feedback to correct the identified tests and guide the iterative process. We use consistency voting from codes to tests (\concodetest{}) to identify tests passed by fewer codes, as these are more likely to contain errors. Correcting the most likely erroneous tests yields greater benefits and reduces the number of iteration rounds. With the support of consistency, the average number of iteration rounds is reduced to just $4$ and the improvement achieves 33\% towards the base model. As the iterations proceed, code quality improves, making the consistency voting increasingly reliable. 

\textit{Secondly, we propose a dynamic strategy to fix and maintain a set of consistent tests and code candidates.} Specifically, \ourtool{} operates iteratively, using a \threestage{} co-evolution process in each iteration to gradually improve the quality of both code and tests. During this co-evolution process, we leverage two forms of consistency voting. The consistency voting from codes to tests (\concodetest) identifies the most likely erroneous tests, as previously mentioned; the consistency voting from tests to codes (\contestcode) selects the code consistent with all tests and, therefore, most likely to be correct—i.e., the code that passes all tests, which serves as the termination condition of the process.  Each iteration of the \threestage{} co-evolution process consists of three stages: (1) we use the consistency voting \concodetest{} to rank the tests and identify the test most likely to be incorrect, (2) users verify the correctness of the tests and make corrections if necessary, and (3) we use the corrected tests to further fix the code using the same \llm{} for generation, then re-rank the tests and proceed to the next iteration. As the co-evolution progresses, the code and tests mutually refine each other, enhancing the quality of both, which makes \concodetest{} and \contestcode{} increasingly reliable. More reliable \contestcode{} allows us to select better code, while  more reliable \concodetest{} enhances the accuracy in identifying incorrect tests. The co-evolution process terminates when we identify a code that passes all tests. Given the high quality of the tests, this selected code is more reliable.

\begin{figure}[t]
    \centering
    \includegraphics[width=0.84\linewidth]{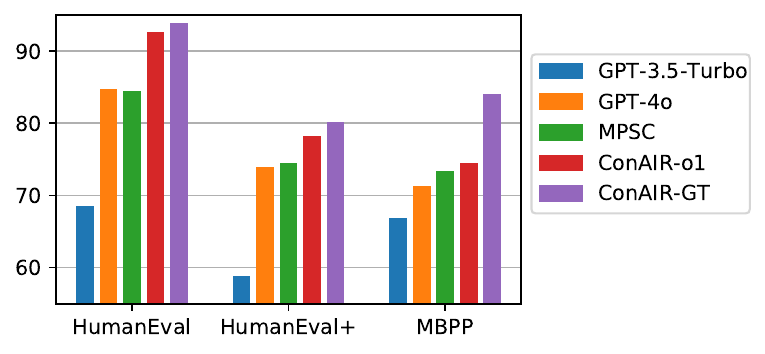}
    \caption{\ourtool{} built on \gptthree{} surpasses the SOTA general \llm{} \gptfour{} and the SOTA post-processing technique \mpsc{} on all datasets.}
    \label{fig:leida}
\end{figure}


We conduct a comprehensive evaluation of \ourtool{}. First, we assess its effectiveness in enhancing \llm{} code generation. With a suboptimal model, \gptthree{}, \ourtool{} achieves an average improvement of 32.9\% over \gptthree{}, an 11.1\% improvement over the state-of-the-art post-processing technique, \mpsc{}\cite{huang2023enhancing}, and a 12.32\% improvement over the most advanced \llm{}, \gptfour{}, as shown in Fig.\ref{fig:leida}. This improvement is achieved with only a 4-round interaction with users, requiring minimal user effort. Additionally, \ourtool{} demonstrates consistent improvements when built on the state-of-the-art \llm{} \gptfour{} (by 16.97\%) and even on the reasoning \llm{} \oo{} (by 8.85\%). We also conduct a user study and examine the overhead of \ourtool{} in terms of time and cost.

In summary, this paper makes the following contributions:
\begin{itemize}[leftmargin=0.5cm]

\item \textbf{A lightweight interaction framework}, which incorporates the user feedback to correct the identified tests and guide the iterative process, resulting in an improvement of 33\% towards the base model and 12\% towards \gptfour{} with only 4 rounds of iteration.

\item \textbf{A \threestage{} co-evolution process} that leverages two forms of consistency voting, with gradually improving code and tests that enhance the reliability of consistency.

\item \textbf{A finding on consistency techniques} that they often overlook preconditions when applying consistency, which can lead to unreliable results.

\item \textbf{A comprehensive evaluation}, which evaluates \ourtool{} from both quantitative experiments and a user study.

\end{itemize}

%% file: sections/motivation.tex

In this section, we introduce the motivation for our work, highlighting that consistency can effectively enhance the reliability of LLM-generated outputs. However, relying solely on consistency is insufficient. First, although the most consistent outputs are more likely to be correct, they may still contain issues. Second, if the \indicator{}s used to assess consistency are of low quality, the resulting consistency will be unreliable.


\subsection{Effectiveness of Consistency and Limitations of Relying Solely on It}

\begin{figure}[t]
    \centering
    \begin{subfigure}{0.6\textwidth}
        \centering
        \includegraphics[width=\linewidth]{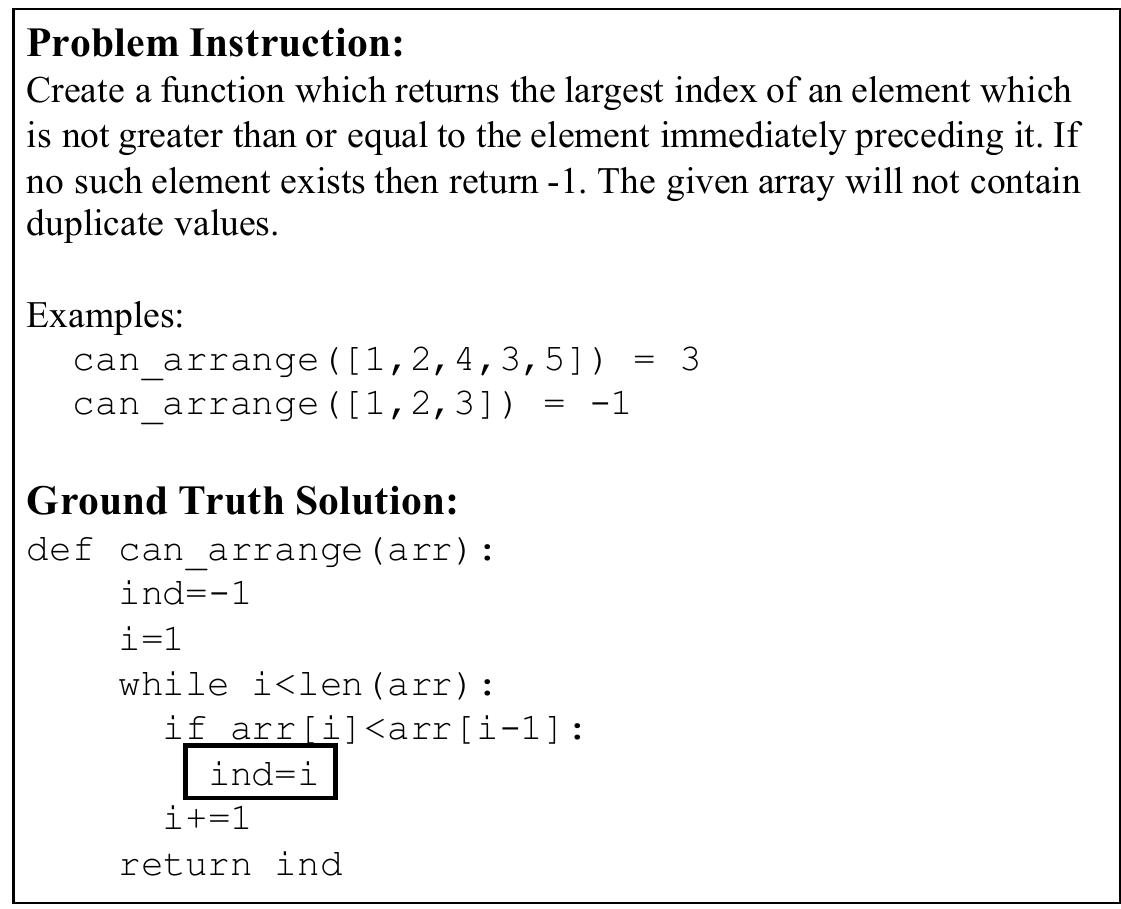} 
        \caption{Problem Instruction}
        \label{fig:moti_fix_intruction}
    \end{subfigure}
    \begin{subfigure}{0.7\textwidth}
        \centering
        \includegraphics[width=\linewidth]{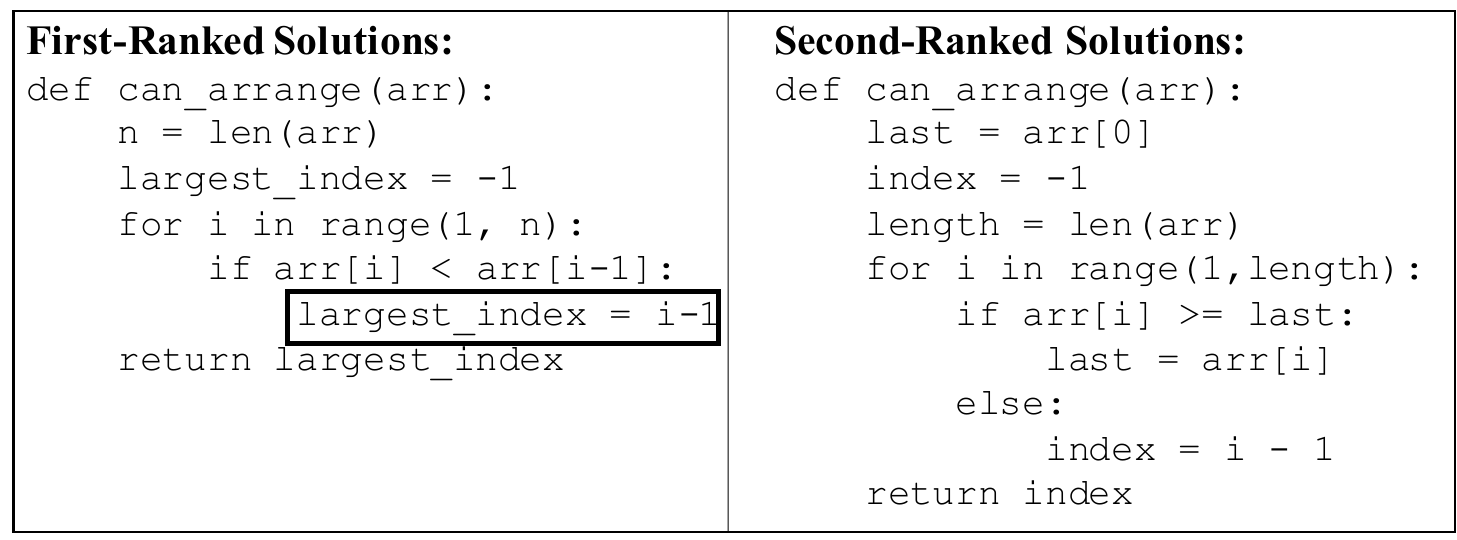} 
        \caption{Top-Ranked and Second-Ranked Solutions}
        \label{fig:moti_fix_solutions}
    \end{subfigure}
    \caption{The motivating example for limitations of relying solely on consistency (\humaneval{}/135)}
    \label{fig:motivating_example_fix}
\end{figure}


Firstly, we discuss the effectiveness of consistency and the limitations of relying solely on it. In Fig.~\ref{fig:moti_fix_intruction}, we present the instructions for \humaneval{}/135, that is, ``finding the largest index of an element that is less than the previous element.'' In Fig.~\ref{fig:moti_fix_solutions}, we display the solutions generated by \llm{}s. The left side represents the solution set with the highest consistency. This set contains the most functionally equivalent counterparts and passes the majority of tests. On the right are other, less consistent solutions. The first-ranked solution on the left is very close to the ground truth solution; its overall logic is correct, but the selected index is slightly off. In contrast, the solution on the right is algorithmically incorrect. This demonstrates the effectiveness of consistency in helping us identify solutions that are more likely to be correct.

Although consistency can increase confidence in selecting correct solutions, relying solely on it is insufficient. While the first-ranked solutions are close to correct, none are entirely accurate; in fact, all generated code for this problem is incorrect. Therefore, beyond consistency, additional post-processing of the generated code is necessary. All existing consistency-based techniques lack post-processing steps and simply select a single output. In this paper, we focus on fixing the code using tests corrected with user feedback. In this example, after just two correction steps, we obtain accurate and consistent codes. Notably, 66.7\% of the final correct solutions originate from the initially top-ranked group, as nearly correct code is easier to fix than entirely incorrect code.

\subsection{Consequences of Neglecting Consistency Preconditions}

\begin{figure}[t]
    \centering
    \begin{subfigure}{0.49\textwidth}
        \centering
        \includegraphics[width=\linewidth]{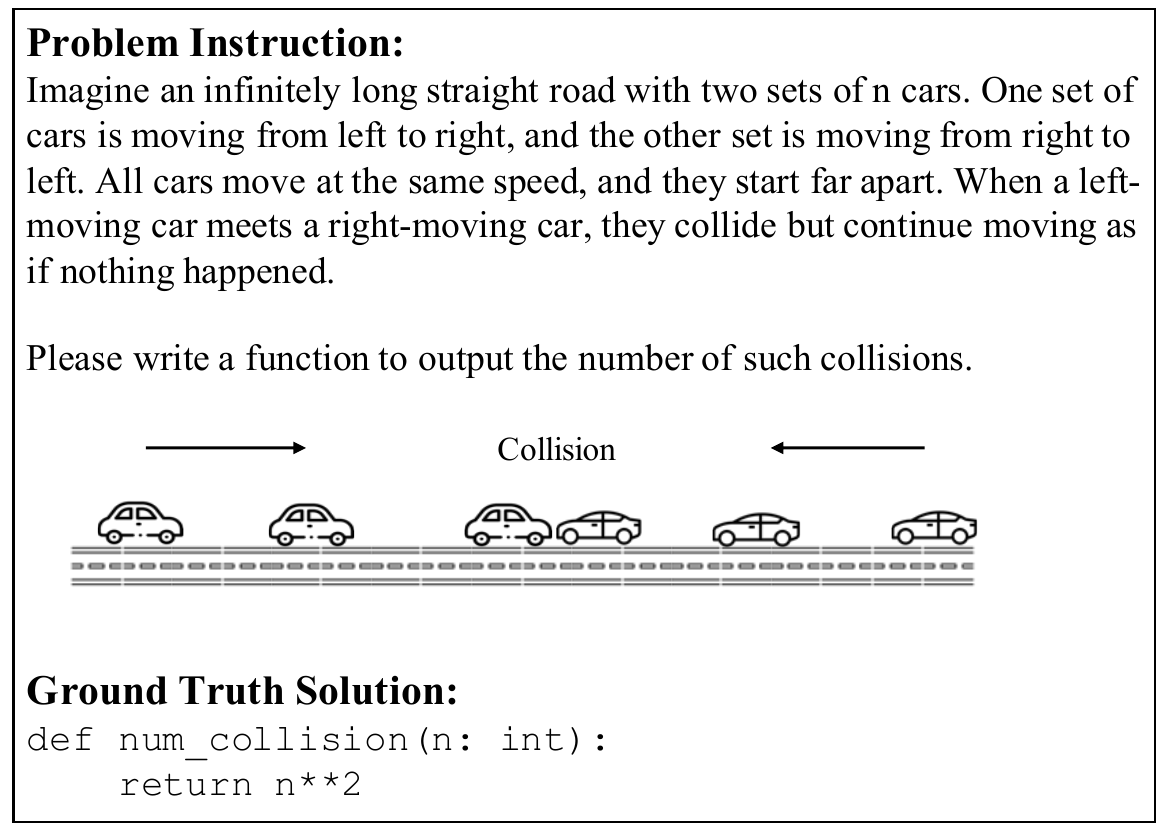} 
        \caption{Problem Instruction}
        \label{fig:moti_instruction}
    \end{subfigure}
    \hfill
    \begin{subfigure}{0.49\textwidth}
        \centering
        \includegraphics[width=\linewidth]{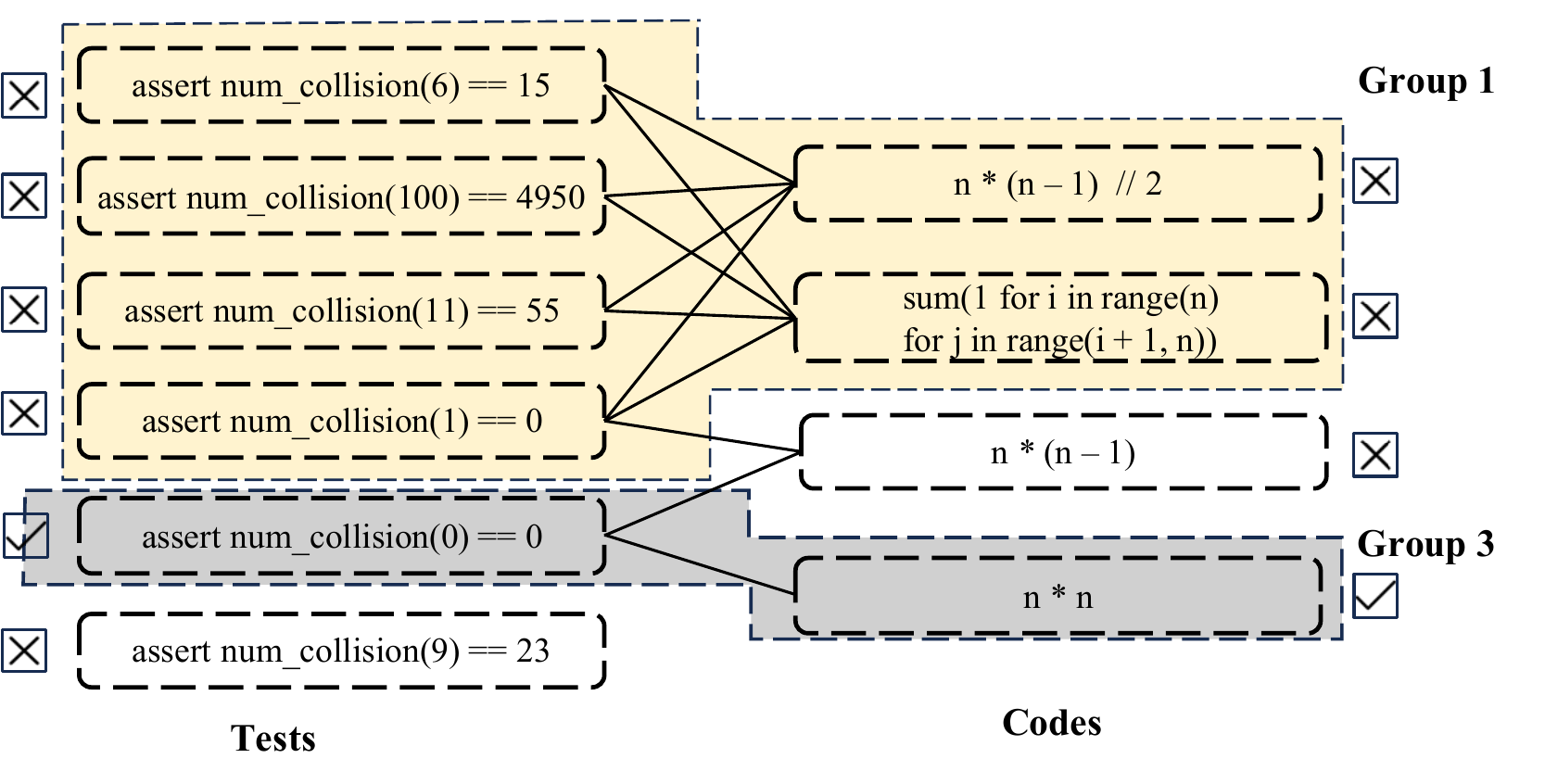} 
        \caption{Testing Results}
        \label{fig:moti_exec}
    \end{subfigure}
    \caption{The motivating example illustrating the consequences of ignoring the \pre{} required for consistency (\humaneval{}/41)}
    \label{fig:motivating_example}
\end{figure}

In addition to the need for further processing of outputs obtained through consistency, there are instances where consistency leads to incorrect results when the \indicator{} (in this case, tests) are of low quality. Existing techniques frequently overlook these prerequisites for effectively utilizing consistency. Specifically, the assessment for consistency relies on relatively good-quality \indicator{}s; without this, consistency achieved through inaccurate \indicator{}s is unreliable, leading to potentially incorrect outputs. To illustrate, consider an example from \humaneval{}~\cite{chen2021evaluating}, as shown in Fig.~\ref{fig:motivating_example} (\humaneval{}/41). The left side (Fig.~\ref{fig:moti_instruction}) presents the problem description and a ground truth solution, while the right side (Fig.~\ref{fig:moti_exec}) shows simplified testing results. In this case, the top-ranked group, containing 34 tests and 44 codes (simplified as Group 1 in Fig.~\ref{fig:moti_exec}), ranks highest by consistency, while a group of 2 tests and 2 codes (simplified as Group 3) ranks lowest. To simplify the graph, we omit Group 2, which consists of the 3-rd code and the 4-th and 5-th tests. The highest-ranked group passes the most tests and contains the most functionality-equivalent \samecode{}s, selected as the final output by existing techniques\cite{chen2022codet,huang2023enhancing}. However, all codes in this group are incorrect, while the correct codes are ranked last. This mistake arises because most tests passed by the top group (32 out of 34) are incorrect. Fig.~\ref{fig:moti_exec} illustrates this with a simplified testing results: here, Group 1 of 4 tests and 2 codes are all incorrect, while the correct code \texttt{``n ** 2''} only passes one test. Therefore, when test quality is low, majority voting based on testing results is unreliable, a limitation overlooked by current methods. Existing techniques\cite{huang2023enhancing,chen2022codet} that rely solely on consistency are unable to select correct answers due to the lowest consistency level of the correct answers. Thus, incorporating user feedback is essential to enhance test quality and reliability.


We leverage consistency voting from codes to tests \concodetest{} to identify the most inconsistent test—that is, the test that most codes fail to pass. In this example, the identified test is the sixth test in Fig.~\ref{fig:moti_exec}. We then prompt users to correct the output, which they adjust from ``23'' to ``81'' based on the instruction requirements. With this corrected test, we can identify the codes that fail it and fix these codes. Here, the initially top-ranked codes fail this test, while the lowest-ranked codes pass. After this fixing, codes that cannot be fixed are discarded. With just one step of user feedback and code fix, all remaining codes align perfectly, producing consistent outputs across all tests. \ourtool{} demonstrates promise by achieving correct results with only minimal human interaction.

%% file: sections/approach.tex
\begin{figure}[t]
    \centering
    \includegraphics[width=0.8\linewidth]{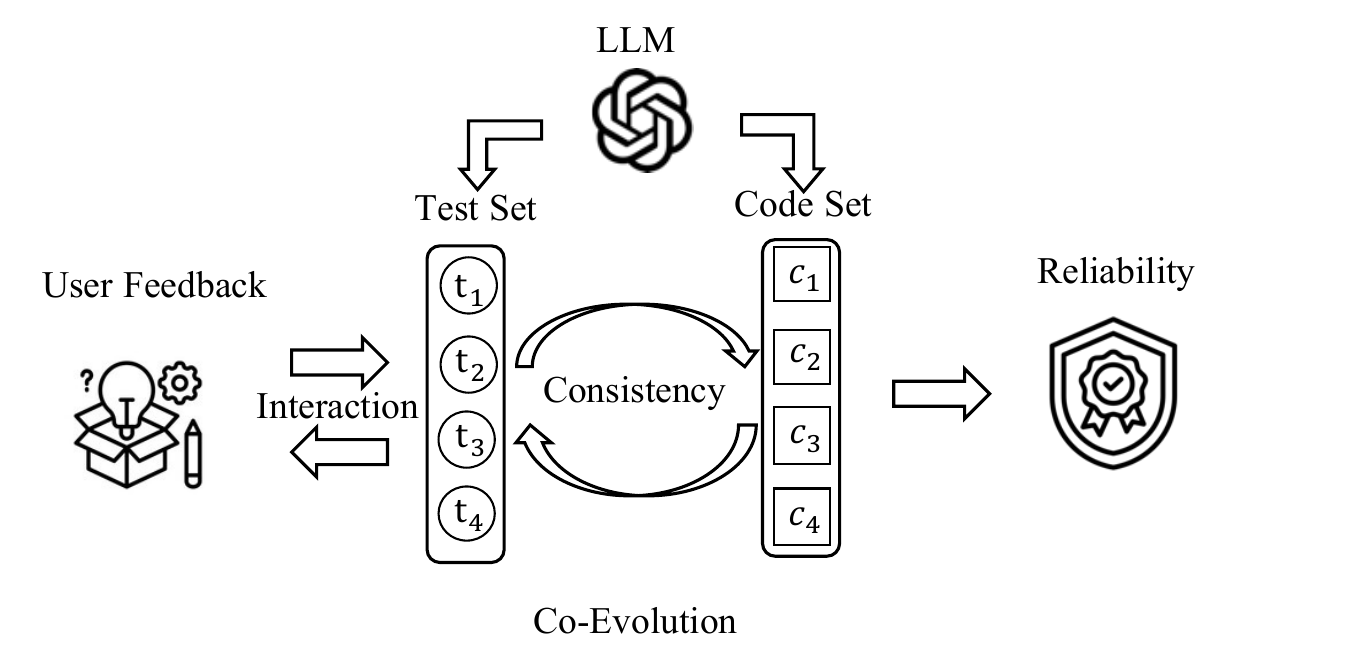}
    \caption{The overview of \ourtool{}}
    \label{fig:overview}
\end{figure}

In this paper, we introduce a new consistency-aided technique, \ourtool{}, which incorporates a lightweight interaction framework to gather user feedback and a co-evolution process to iteratively enhance the quality of both tests and codes.  An overview of the workflow of \ourtool{} is illustrated in Fig.~\ref{fig:overview}. \ourtool{} has two ingredients that distinguish it from existing approaches. Firstly, the developers are involved as the ultimate oracle, and we propose a lightweight interaction framework that incorporates user feedback to correct the identified tests and guide the iterative process. Secondly, \ourtool{} uses a \threestage{} co-evolution process in each iteration to gradually improve the quality of both code and tests, which makes \concodetest{} and \contestcode{} increasingly reliable. In the end, we achieve a more reliable code through improved consistency.
In the following, we introduce the details of \ourtool{}. Specifically, we will introduce the task definition in Section~\ref{subsec:task_definition}, the overall interaction framework in Section~\ref{subsec:interaction}, and the specific co-evolution process in each iteration in Section~\ref{subsec:coevolution}.

\subsection{Task Definition}
\label{subsec:task_definition}

We first introduce the definition and setup of this task, that is, leveraging consistency to improve LLM code generation results. The code generation task aims to generate a code solution, \code{}, based on a problem description, \des{}, using a large language model, \model{}. Formally, this is represented as \code{} = \model{}(\des{}). The problem description, \des{}, provides the requirements in natural language and includes the function signature, specifying the function name and parameters, as shown in the example in Fig~\ref{fig:moti_fix_intruction}. Generating correct code in a single attempt is challenging for \llm{}s~\cite{chen2022codet,huang2023enhancing}. To address this, researchers propose sampling multiple code solutions from \llm{}s, denoted as \codeset{} = \{\code{}$_1$, \code{}$_2$, ..., \code{}$_n$\}, and obtaining a code \bestcode{} based on \codeset{} that is more likely to be correct. In addition to generating code, researchers also use the same \llm{}, \model{}, to generate a set of tests, \testset{} = \{\test{}$_1$, \test{}$_2$, ..., \test{}$_m$\}, to aid in obtaining the best code, \bestcode{}. These tests, \testset{}, serve as \indicator{}s to evaluate the consistency of the generated codes, \codeset{}.
A test case \test{} is defined as a pair of input and expected output (i.e., \test{} = $(x, y)$), which verifies whether the output of the code \code{} meets the requirements specified in the problem description \des{}. Existing consistency-based LLM code generation methods~\cite{chen2022codet,huang2023enhancing} work by selecting the code that passes the most tests (inter-consistency) and has the highest number of functionally equivalent \samecode{}s (intra-consistency). However, this approach can be problematic: while tests help verify code correctness, they may also be incorrect since they are generated by the same \llm{}. In this paper, we introduce a new consistency-augmented technique, \ourtool{}, which incorporates an interaction framework to gather user feedback \feedbackset = \{\feedback{}$_1$, \feedback{}$_2$, ..., \feedback{}$_k$\} and a co-evolution process to iteratively enhance the quality of both tests and codes. In our approach, the input and output are defined as \bestcode{} = \ourtool{}(\codeset{}, \testset{}, \feedbackset{}).

\subsection{Lightweight Interaction Framework to Gather User Feedback}
\label{subsec:interaction}
In this section, we present the overall framework of \ourtool{}. \ourtool{} is a lightweight interaction framework that collects user feedback to correct identified tests and guide the iterative improvement process. To satisfy the \pre{} necessary for consistency, we incorporate user feedback in a developer-friendly, lightweight manner. In our task, tests act as \indicator{}s that verify code correctness and assess consistency. However, our experiments reveal that 37.7\% of tests generated by the LLM are incorrect. With only the tests \testset{} and the codes \codeset{}, we cannot ensure test accuracy or determine their correct outputs, as both tests and codes are generated by the same \llm{}, making them potentially unreliable. Additionally, in many cases, only the developer can determine the correctness of specific test results. In real-world development, users must ensure the correctness
of tests; otherwise, the quality of the code may be compromise. Keeping the developer in the loop also fosters a sense of code ownership and ensures that they maintain a clear understanding of the code generation process, enabling them to respond more swiftly to future bugs. Therefore, integrating user feedback is essential—a need also supported by many studies~\cite{norman2013design,christiano2017deep}.

However, it is crucial to minimize the need for developer consultation. First, we aim to simplify the questions users need to answer. Therefore, \ourtool{} prompts users to check and correct the tests rather than directly fixing the code. Second, we limit the user check to the fewest possible rounds. To achieve this, we use consistency voting from codes to tests (\concodetest{}) to identify tests that are passed by fewer codes, as these are more likely to contain errors. Following existing work~\cite{chen2022codet,huang2023enhancing}, we formalize the consistency relationship between code \code{} and test \test{} as 

\begin{equation}    
\texttt{Con}(c, t) =
\texttt{Con}(c, (x,y)) = 
\left\{
\begin{aligned}
    &True, c(x) = y \\
    &False, c(x) \neq y
\end{aligned}
\right.
\end{equation}
Tests and codes are implementations of the same problem requirements from two different perspectives. The code \code{} and the test \test{} are considered consistent when \code{} passes \test{}, indicating that the functionality aligns from both perspectives. The consistency voting \concodetest{} represents the degree of consistency between each test \test{} and the entire code set \codeset{}, serving as a measure of the test’s reliability from the perspective of the codes and is denoted as
\begin{equation}
\concodetesttext{}(\testtext{}, \codesettext{}) = \sum_{\codetext{}}{\context(\testtext{}, \codetext{})}
\end{equation}
The lower the consistency voting (\concodetest{}), the more likely it is that the test is incorrect. We rank the tests based on \concodetest{} and select the most inconsistent test for correction, which can enhance the quality of both the test set and code set and thus reduces the number of iteration rounds. Additionally, when the test is consistent with all codes—i.e., $\concodetesttext{}(\testtext{}, \codesettext{}) = \text{size}(\codesettext{})$—we skip the correction process and directly use the code outputs as the test output. Although individual codes may be incorrect, the likelihood of all codes producing the same incorrect result is low. In addition, as iterations proceed, code quality steadily improves, making consistency voting increasingly reliable. By leveraging these two methods, we decrease the number of iteration rounds and reduce the need for human feedback. With the support of consistency, the average number of iteration rounds is reduced to just 4, with an improvement of 33\% to the base model.


\begin{figure}[t]
    \centering
    \includegraphics[width=\linewidth]{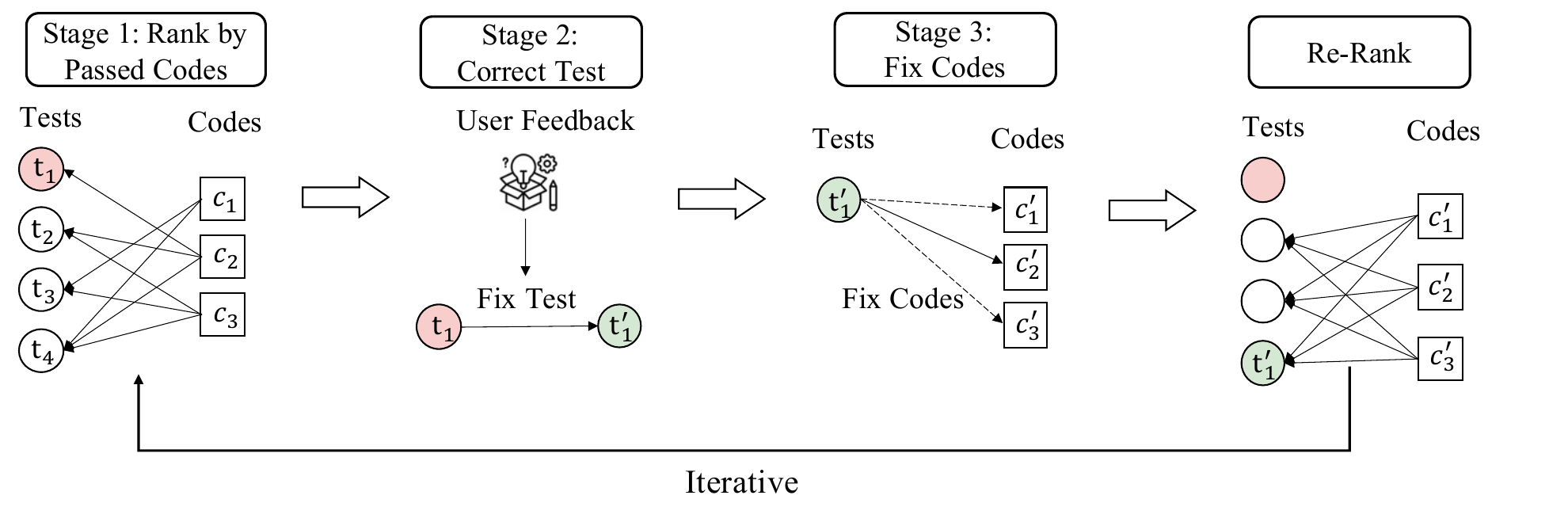}
    \caption{The co-evolution process of \ourtool{}}
    \label{fig:coevolution}
\end{figure}

\subsection{Co-Evolution Process between Codes and Tests}
\label{subsec:coevolution}

In Section~\ref{subsec:interaction}, we introduced the overall interaction framework. Here, we detail the specific process within each iteration. Each iteration leverages a \threestage{} co-evolution process between codes and tests to iteratively improve their quality, as illustrated in Fig.\ref{fig:coevolution}. The co-evolution algorithm is outlined in Algorithm~\ref{alg:correct_code}.

During the co-evolution process, we employ two types of consistency voting. The consistency voting from codes to tests (\concodetest) identifies tests most likely to be erroneous, as discussed in Section~\ref{subsec:interaction}. The consistency voting from tests to codes (\contestcode) selects the code consistent with all tests and, therefore, most likely to be correct—i.e., the code that passes all tests, serving as the process's termination condition. Through this interaction, codes and tests co-evolve, enhancing each other. The tests help identify erroneous cases, and once corrected, they further aid in refining the codes. As the quality of both codes and tests improves, the reliability of \concodetest{} and \contestcode{} increases. Higher-quality codes make it more likely that tests they cannot pass are buggy, while improved tests increase the likelihood that codes passing more tests are correct.

\begin{algorithm}
    \small
    \SetAlgoNlRelativeSize{-1}
    \KwIn{test case set $\mathbf C$; code set $\mathbf T$}
    $\mathbf T_{\text{unk}} \leftarrow \mathbf T$;\ \ 
    $\mathbf T_{\text{cor}} \leftarrow \{\}$;\ \ 
    $\mathbf C_{\text{dis}} \leftarrow \{\}$\Comment*[r]{initializing sets}
    
    
    \While{true}{
        $t_{\text{w}} \leftarrow \argmax_{t \in \mathbf T_{\text{unk}}} \texttt{Con}_{c \to t}(t, \mathbf C)$ \Comment*[r]{rank and localize a worst test case}
        
        $t_{\text{cor}} \leftarrow \mathrm{InteractivelyCorrectTestCase}(t_{\text{w}})$ \Comment*[r]{interactively correct the worst test case}
        
        $\mathbf T_{\text{unk}}.\mathrm{remove}(t_{\text{w}})$;\ \ 
        $\mathbf T_{\text{cor}}.\mathrm{add}(t_{\text{cor}})$;
        
        $\mathbf C_{\text{rem}} \leftarrow \{\}$;
        
        \For{$c \in \mathbf C$}{
            \If{$\mathtt{Con}(t_{\text{cor}}, c)$}{
                $\mathbf C_{\text{rem}}.\mathrm{add}(c)$;
            }
            \Else{
                $c' \leftarrow \mathrm{LLMFixCode}(c)$ \Comment*[r]{fix the code by LLM}
                
                \If{$\mathtt{Con}_{t \to c}(\mathbf T_{\text{cor}}, c') = \mathrm{size}(\mathbf T_{\text{cor}})$}{
                    $\mathbf C_{\text{rem}}.\mathrm{add}(c')$;
                \Comment*[r]{the fixed code can pass all the corrected tests}}
                \Else{
                    $\mathbf C_{\text{dis}}.\mathrm{add}(c')$;
                }
            }
        }
        \If(\Comment*[f]{if no code passes corrected test cases}){$\mathbf C_{\text{rem}}\mathrm{.isEmpty()}$}{
            \Return $\argmax_{c \in \mathbf C_{\text{dis}}} \mathtt{Con}_{t \to c}(\mathbf T_{\text{unk}} \cup \mathbf T_{\text{cor}}, c)$;
        }
        $\mathbf C \leftarrow \mathbf C_{\text{rem}}$ \;
        
        \If(\Comment*[f]{if all test cases are corrected}){$\mathbf T_{\text{unk}}\mathrm{.isEmpty()}$}{
            \Return $\argmax_{c \in \mathbf C} \mathtt{Con}_{t \to c}( \mathbf T_{\text{cor}}, c)$;
        }
        
        \For(\Comment*[f]{find a code that passes all test cases}){$c \in \mathbf C$}{
            \If{$\mathtt{Con}_{t \to c}(\mathbf T_{\text{unk}} \cup \mathbf T_{\text{cor}}, c) = \mathrm{size}(\mathbf T_{\text{unk}} \cup \mathbf T_{\text{cor}})$}{
                \Return $c$;
            }
        }
    }
\caption{Co-Evolution Algorithm between Codes and Tests}
\label{alg:correct_code}
\end{algorithm}


Next, I will provide a detailed introduction to the co-evolution algorithm, using Algorithm~\ref{alg:correct_code} and Fig.~\ref{fig:coevolution} to offer a more intuitive explanation. As shown in Algorithm~\ref{alg:correct_code}, the input consists of the test set \testset{} and code set \codeset{} generated by the \llm{} \model{}, with the output being a more reliable and accurate code, \bestcode{}, than any individual code in \codeset{}. Before the algorithm begins, each code is executed on each test to gather execution information. This process is efficient, as it can be parallelized.
The main body of the algorithm consists of iterations, where tests are divided into two groups: \textit{corrected tests}, \cortestset{}, and \textit{unknown tests}, \unktestset{}. The iteration terminates once all unknown tests are corrected or a code that passes all tests is found. Each iteration involves a \threestage{} co-evolution process between codes and tests, divided into three stages: ``ranking'', ``correcting'', and ``fixing''.


\paragraph{Ranking} The first stage, "ranking," is outlined in Line 3 of Algorithm~\ref{alg:correct_code}. As already introduced in Section~\ref{subsec:interaction}, we leverage the consistency voting from codes to tests (i.e., \concodetest{}) to rank the tests and identify the test that is most likely to be incorrect, denoted as \wrongtest{}. As the co-evolution process iterates, the quality of codes improves, leading to increasingly accurate identification. As shown in Stage 1 of Fig.~\ref{fig:coevolution}, $\testtext{}_4$ has three codes that pass it, $\testtext{}_3$ and $\testtext{}_2$ have two codes that pass each, and $\testtext{}_1$ only has one code that passes it. Based on the consistency voting from codes to tests, \concodetest{}, $\testtext{}_1$ is most inconsistent with the codes, making it more likely to contain an error and thus selected.


\paragraph{Correcting.} In Line 4-5, describes the second stage, "correcting", where we utilize an interaction process to incorporate user feedback, verifying the accuracy of the identified test \wrongtest{} and correcting it if necessary. Once corrected, \wrongtest{} is updated to \cortest{}. As shown in Stage 2 of Fig.~\ref{fig:coevolution}, $\testtext{}_1$ is updated to $\testtext{}_1^{'}$. This test is then removed from the unknown set \unktestset{} and added to the corrected set \cortestset{}. Once corrected, we also update the consistency relationships between tests and code. This stage produces the first update of the consistency voting results, making the consistency voting from tests to code \contestcode{} more reliable.


\paragraph{Fixing.} The third stage, ``fixing'', described in Lines 7-15 of Algorithm~\ref{alg:correct_code}, involves fixing the codes that fail on the corrected test \cortest{}. For this, we employ the same model \model{} that initially generated codes to fix the codes, allowing us to demonstrate the improvements brought by \ourtool{}. As shown in Stage 3 of Fig.~\ref{fig:coevolution}, $c_1$, $c_2$, and $c_3$ initially fail to pass $t_1^{'}$. We leverage \model{} to fix them, resulting in the corrected versions $c_1^{'}$, $c_2^{'}$, and $c_3^{'}$.

After fixing the code, we check whether the corrected code can pass all tests in the corrected set \cortestset{} (Line 12). We divide the code set \codeset{} into \textit{discarded codes} \discodeset{} and \textit{retained codes} \remcodeset{}. If the fixed code \fixcode{} cannot pass \cortestset{}, it is removed from \remcodeset{} and added to \discodeset{}. This process highlights the advantage of introducing diversity in the generated codes. Some implementations may contain fundamental logical errors that are difficult to fix, while others may only fail on certain edge cases, making them fixable. Diversity thus enhances the chances of obtaining a viable code solution. If all codes end up in \discodeset{}, indicating the absence of a code that fully satisfies all tests, we select and output the code from \discodeset{} that passes the most tests (Line 16-17).


After fixing all codes, we execute the fixed code \fixcode{} on the unknown test set \unktestset{} and update the consistency relationships to re-rank the tests. This produces the second update of the consistency relationship. After updating, if we identify a code that passes all tests, we output this code as \bestcode{} (Line 21-23). With the improved quality of tests, consistency voting from tests to code becomes increasingly reliable, meaning a code that passes all tests is likely to be correct. If no such code exists, we proceed to the next iteration. Enhanced code quality strengthens consistency, allowing for more precise identification of incorrect tests in subsequent iterations. As shown in the ``re-rank'' section of Fig.~\ref{fig:coevolution}, $t_3$ has no passing codes, while other tests are passed by all codes. Therefore, $t_3$ is selected for further inspection, initiating the next iteration.

%% file: sections/setup.tex
\subsection{Research Questions}
We begin by evaluating the performance of \ourtool{} through two automated simulated experiments: one where user feedback is simulated using \llm{} \oo{}, which has advanced reasoning capabilities, and another where feedback is simulated via the \gtsolution{}. Following these, we examine the user efforts required to use \ourtool{} and conduct a user study to assess the user experience in a real interactive setting. Lastly, we analyze the overhead of \ourtool{}.


\begin{itemize}[leftmargin=0.5cm]
\item \textbf{RQ1. Overall Effectiveness} How effective is \ourtool in improving LLM code generation? We conduct two simulated experiments to automatically evaluate \ourtool{}, enabling extensive quantitative analysis. In the first experiment, \oo{} is used to simulate providing user feedback, while in the second experiment, the \gtsolution{} is used to simulate providing user feedback.
\item \textbf{RQ2. User Efforts and User Study} What is the user effort required by \ourtool? We provide and discuss the user effort required by \ourtool{} in the experiments. We further report the user study here, which assesses the user experience in a real interactive setting.
\item \textbf{RQ3. Time and Cost Overhead} What is the overhead of \ourtool? We analysis the time and cost of \ourtool{}.
\end{itemize}

\subsection{Dataset}
We adopt three standard code generation datasets that are widely-used by code generation techniques~\cite{luo2023wizardcoder,achiam2023gpt,guo2024deepseek,nijkamp2022codegen,du2021glm,li2023starcoder,fried2022incoder, hui2024qwen2}. \humaneval~\cite{chen2021evaluating} and \mbpp{}~\cite{austin2021program} comprise a set of hand-crafted Python programming problems. For each problem, the dataset provides the problem description in natural language, the tests to check the correctness of given output, and the ground truth solution. \humanevalplus{}~\cite{liu2024your} adds more tests to the \humaneval{} dataset, which makes the check more strict. The statistics of the three datasets are as follows. The number of problems in \humaneval{}, \humanevalplus{}, and \mbpp{} are $164, 164, 427$ respectively. The average number of tests in three datasets are $7.77, 764.74, 3.1$ respectively. 
\subsection{Evaluation Metrics}
Following all the code generation techniques, we use the \passk{} metric. For each problem, we may generate multiple code solutions and select the top $k$ as the final candidates. Among these $k$ solutions, if at least one successfully passes all the tests for the problem, the problem is considered solved. \passk{} represents the ratio of successfully solved problems to the total number of problems.
\subsection{Compared Techniques}
Firstly, we compare \ourtool{} with \openai{} models, including \gptthree{}, and the most advanced general model, \gptfour{}. Additionally, we compare with other \llm{}s for code generation, such as Code Llama~\cite{roziere2023code}, WizardCode~\cite{luo2023wizardcoder}, and Deepseek Code~\cite{guo2024deepseek}. Since \ourtool{} functions as a post-processing technique for outputs generated by \llm{}, we also evaluate it against other post-processing methods, including the state-of-the-art \mpsc{}\cite{huang2023enhancing} and \codet{}\cite{chen2022codet}. These two techniques are also based on consistency. \codet{} uses tests as \indicator{}s, selecting the code that satisfies the most tests and has the highest number of \functionequal{} \samecode{}s based on test results. \mpsc{} further incorporates \spec{}s as additional \indicator{}s; however, the benefits brought by \spec{} are limited. Other post-processing techniques in the comparison include Self-Consistency~\cite{wang2022self}, MBR-EXEC~\cite{shi2022natural}, and Self-Collaboration~\cite{dong2024self}.

\subsection{Implementation}
In our experiments, we choose \gptthree{}, \gptfour{} and \oo{} as the base foundation model. We leverage the same model to generate codes and tests, and also fix the codes. \ourtool{} achieves a consistent improvement on three base models. To keep consistent with the \sota{} technique \mpsc{}, the \gptthree{} version we adopt is gpt-3.5-turbo-0613 API, which is released on 2023-06-13. \gptthree{} is an old and less powerful model, and the context size is only 4,096. The \gptfour{} version we adopt is gpt-4o-2024-08-06, which is the latest and most powerful general \llm{}. The context size is 128,000. \oo{}~\cite{o1} is the latest \llm{} released by \openai{}, noted for its strong reasoning abilities and its use of chain-of-thought processes to enhance generation quality. \ourtool{} can achieve consistent improvement on both poor and powerful model, which indicates the effectiveness of \ourtool{}. The performance of \ourtool{} based on \gptthree{} is even better than \gptfour{}. In RQ1 (Section~\ref{sec:overall}), we also investigate scenarios where user feedback is simulated with \oo{}. \oo{} can solve the reasoning and logic problems with chain-of-thought, which can simulate humans. The version we use is o1-preview-2024-09-12.

%% file: sections/results.tex

In this section, we present the experimental results along with an in-depth analysis. First, we conduct two simulated experiments to automatically evaluate \ourtool{}’s effectiveness, as detailed in Section~\ref{sec:overall}. Next, we discuss the user effort involved in using \ourtool{} during these experiments and include a user study assessing user experience in a real interactive setting in Section~\ref{sec:user_study}. Finally, we analyze the time and cost overhead associated with \ourtool{} in Section~\ref{sec:cost}.

\input{sections/rq1}
\input{sections/rq2}

\input{sections/rq4}

%% file: sections/rq1.tex
\subsection{RQ1: Overall Effectiveness}
\label{sec:overall}

\subsubsection{Experimental Design}

In this paper, we propose a lightweight interaction framework to gather user feedback, which enhances the quality of \indicator{} by asking users to validate the correctness of selected tests. In this research question (RQ1), we investigate the effectiveness of \ourtool{} through large-scale, automated experiments. These experiments simulate user feedback, enabling extensive quantitative analysis. In Section~\ref{sec:user_study} (RQ2), we conduct a user study to evaluate \ourtool{} in a real interactive setting. Given the limitations in scaling the user study to a large scale, we leverage simulated user interactions in RQ1. Specifically, we use two simulation methods: (1) \oo{} is used to simulate providing user feedback, (2) \gtsolution{} is used to simulate providing user feedback, which can be regarded as \textit{novice users} and \textit{experienced users} respectively. Accordingly, the two variants of \ourtool{} are referred to as \ourtooloo{} and \ourtoolgt{}, respectively.

Experienced users rarely make mistakes, while novice users are more prone to errors. In real development scenarios, tests should be error-free, as they serve to ensure code correctness. They are typically provided by users, who are responsible for ensuring their accuracy. If tests contain errors, the quality of the generated code is compromised. Additionally, it is often easier for users to provide expected test outputs rather than writing code, as they already have a clear understanding of the requirements. In simulations using \oo{}, we employ \oo{} to correct the identified tests. This process is fully automated, requiring no user involvement. Therefore, \ourtooloo is an automated variant of \ourtool. In simulations using \gtsolution{}, we utilize the \gtsolution{} provided by the benchmark to execute the identified tests and generate ground truth outputs as user feedback.
 
For the base model, we ensure fair comparison by following the setup in \sotasimple{}\cite{huang2023enhancing}, using \gptthree{} for code generation and fixing, as shown in Table~\ref{tab:overall}. Besides, we also build \ourtool{} based on the most advanced general \llm, \gptfour{}, and the latest reasoning \llm, \oo{}, where \ourtool{} consistently shows performance improvements, as shown in Table~\ref{tab:powerful}. Due to the significant time and cost associated with \oo{} (detailed in Section~\ref{sec:cost}), we sampled 25 problems from \humaneval{} for code generation and fixing using \oo{}. \oo{} is more suitable for complex logic and math tasks, and its cost makes it impractical for routine code generation.
\input{tables/overall}

\subsubsection{Experimental Results}

In Table~\ref{tab:overall}, we present the effectiveness of \ourtool{} across two simulated experiments. In these experiments, \ourtooloo uses \oo{} to simulate feedback, while \ourtoolgt{} employs \gtsolution{} for feedback simulation. We conduct experiments on three datasets and compare the results with various baselines. These comparison techniques are grouped into three categories: the top part in Table~\ref{tab:overall} leverages only the \llm{}, the middle part applies post-processing techniques, and the bottom part represents our proposed approach. The increases shown to the right of the post-processing technique results indicate their improvements over the base model, \gptthree{}. As shown in Table~\ref{tab:overall}, both variants of \ourtool{} achieve \sotasimple{} performance across all benchmarks, demonstrating the effectiveness of \ourtool{}. \ourtoolgt{} outperforms \ourtooloo{}, as the outputs assigned to tests in \ourtooloo{} may be incorrect, reducing the model's performance. Built on a suboptimal model, \gptthree{}, \ourtool{} achieves an average improvement of 32.9\% over \gptthree{}, an 11.1\% gain over the state-of-the-art post-processing technique, \mpsc{}~\cite{huang2023enhancing}, and a 12.32\% improvement over the most advanced \llm{}, \gptfour{}.

Next, we compare the performance of \ourtoolgt{} and \ourtooloo{}. The results of \ourtooloo{} on \humaneval{} and \humanevalplus{} are similar, indicating that the outputs generated by \oo{} are close to the ground truth outputs. \ourtooloo{} performs worse than \ourtoolgt{} on \mbpp{}, but it still surpasses the \sotasimple{} baseline, \mpsc{}. This difference arises because many of the outputs generated by \oo{} on \mbpp{} are incorrect. We compute the error rate of the outputs generated by \oo{} for tests; the error rate on \humaneval{} and \mbpp{} are 8.6\% and 48.3\%, respectively. Since the generated outputs guide code fixing to meet their specifications, any inaccuracies can degrade code quality. The \humaneval{} dataset includes comprehensive problem instructions, complete function signatures, and parameter and return value types, which aids understanding. Conversely, the \mbpp{} dataset often lacks parameter and return types, and its problem descriptions can be misleading. In addition, there are no input-output examples to clarify requirements. This ambiguity brings challenges even for human understanding, as noted by participants in the user study. For instance, in \mbpp{}/299, the instruction is "Write a function to calculate the maximum aggregate from the list of tuples," which is vague. Given the test input, [(1, 40), (2, 50), (3, 60), (1, 70), (2, 80), (3, 90)], it is particularly challenging to deduce the intended output based solely on the description.

\input{tables/powerful}

Finally, to demonstrate the generalization ability of \ourtool{}, we further implement it using more powerful \llm{}s, \gptfour{} and \oo{}. The results are shown in Table~\ref{tab:powerful}. In the o1-based \ourtool{}, since the code is already generated by \oo{}, we evaluate only \ourtoolgt{}. As shown in Table~\ref{tab:powerful}, \ourtool{} achieves notable gains with these advanced models as the base. For example, \ourtooloo{} based on \gptfour{} increases performance on \humaneval{} from 84.67 to 97.59. Specifically, built on \gptfour{}, the pass@1 improvements of \ourtooloo{} across three datasets are 12.2\%, 11.0\%, and 5.9\%, while \ourtoolgt{} achieves 15.3\%, 15.82\%, and 19.8\% improvements, respectively. When using \oo{} as the base model, \ourtoolgt{} achieves improvements of 9.94\% and 1.77\%. These increases are smaller than those on \gptfour{} due to the higher initial performance of \oo{}, which limits the room for further gains.


%% file: tables/overall.tex
\begin{table}[t]
    \caption{The performance of \ourtool{} and other baselines on three benchmarks. The best and second-best performances for each dataset are highlighted in \textbf{bold} and \underline{underlined}, respectively.}
    \label{tab:overall}
    \centering
    \begin{adjustbox}{width=\columnwidth}
    \begin{tabular}{lllllll}
        \hline
        \textbf{Benchmark} & \multicolumn{3}{c}{\textbf{\humaneval{}}} & \multicolumn{3}{c}{\textbf{\humanevalplus{}}} \\
        \cmidrule(lr){2-4} \cmidrule(lr){5-7}
        \textbf{Metric} & \textbf{Pass@1} & \textbf{Pass@2} & \textbf{Pass@5} & \textbf{Pass@1} & \textbf{Pass@2} & \textbf{Pass@5} \\ 
        \hline
        \gptfour{} &84.67 & 89.69 & 92.50 &73.82 & 81.03 & 85.80\\
        GPT-3.5-Turbo & 68.38 & 76.24 & 83.15 & 58.75 & 66.58 & 73.96 \\
        DeepSeekCoder & 79.30 &  -&  -&  -&  -& -  \\
        WizardCoder & 73.20 &  -&  -&  -&  -& - \\
        Code Llama & 62.20 &  -&  -&  -&  -& - \\
        \hline
        Self-consistency & 73.86\textsubscript{+5.48} & 73.93\textsubscript{-2.31} & 74.10\textsubscript{-9.05} & 63.50\textsubscript{+4.75} & 64.70\textsubscript{-1.88} & 65.67\textsubscript{-8.29} \\
        MBR-EXEC & 72.96\textsubscript{+4.58} & 76.47\textsubscript{+0.23} & 79.00\textsubscript{-0.45} & 62.12\textsubscript{+3.37} & 67.08\textsubscript{+0.50} & 71.38\textsubscript{-2.58} \\
        CodeT & 78.05\textsubscript{+9.67} & 78.05\textsubscript{+1.81} & 78.30\textsubscript{-4.85} & 67.87\textsubscript{+9.12} & 68.75\textsubscript{+2.17} & 69.65\textsubscript{-4.31} \\
        Self-collaboration & 74.40\textsubscript{+6.02} &  -& -  & - & - & - \\
        MPSC & 84.29\textsubscript{+15.91} & 86.79\textsubscript{+10.55} & 87.13\textsubscript{+3.98} & 74.39\textsubscript{+15.64} & 76.66\textsubscript{+10.08} & 77.25\textsubscript{+3.29} \\
        \hline
        \ourtooloo{} (GPT-3.5-Based) & \underline{92.45\textsubscript{+24.07}} & \underline{94.45\textsubscript{+18.21}} & \underline{95.71\textsubscript{+12.56}} & \underline{78.03\textsubscript{+19.28}} & \underline{82.65\textsubscript{+16.07}} & \underline{86.35\textsubscript{+12.39}}\\
        \ourtoolgt{}  (GPT-3.5-Based) &\textbf{93.71\textsubscript{+25.33}} & \textbf{94.86\textsubscript{+18.62}} & \textbf{96.32\textsubscript{+13.17}} & \textbf{80.00\textsubscript{+21.25}} & \textbf{84.67\textsubscript{+18.09}} & \textbf{88.42\textsubscript{+14.46}}\\
        \hline
        \textbf{Benchmark} & \multicolumn{3}{c}{\textbf{\mbpp}} & & & \\
        \cmidrule(lr){2-4}
        \textbf{Metric} & \textbf{Pass@1} & \textbf{Pass@2} & \textbf{Pass@5} & & & \\
        \cline{1-4}
        \gptfour{} & 71.19 & \underline{76.49} & \underline{79.89} &&&\\
        GPT-3.5-Turbo & 66.80 & 72.34 & 76.60 & & & \\
        DeepSeekCoder & 70.00 & -& -& & & \\
        WizardCoder & 61.20 & -& -& & & \\
        Code Llama & 61.20 & -& -& & & \\
        \cline{1-4}
        Self-consistency & 71.70\textsubscript{+4.90} & 71.73\textsubscript{-0.61} & 71.82\textsubscript{-4.78} & & & \\
        MBR-EXEC & 70.79\textsubscript{+3.99} & 73.14\textsubscript{+0.80} & 74.85\textsubscript{-1.75} & & & \\
        CodeT & 71.90\textsubscript{+5.10} & 71.95\textsubscript{-0.39} & 72.02\textsubscript{-4.58} & & & \\
        Self-collaboration & 68.20\textsubscript{+1.40} & -& -& & & \\
        MPSC & 73.23\textsubscript{+6.43} & 73.29\textsubscript{+0.95} & 73.55\textsubscript{-3.50} & & & \\
        \cline{1-4}
        \ourtooloo{} (GPT-3.5-Based) & \underline{74.30\textsubscript{+7.50}} & 74.53\textsubscript{+2.19} & 75.14\textsubscript{--1.46} \\
        \ourtoolgt{} (GPT-3.5-Based) & \textbf{83.90\textsubscript{+17.10}} & \textbf{86.80\textsubscript{+14.46}} & \textbf{87.45\textsubscript{+10.85}} \\
        \cline{1-4}
    \end{tabular}
    \end{adjustbox}

\end{table}

%% file: tables/powerful.tex


\begin{table}[t]
\caption{The performance of \ourtool{} built on \gptfour{} and \oo{}}
\label{tab:powerful}
\centering
\begin{adjustbox}{width=\columnwidth}
\begin{tabular}{l|ccc|ccc|ccc}
\hline
\multirow{2}{*}{\textbf{Models}} &\multicolumn{3}{c|}{\textbf{HumanEval}} & \multicolumn{3}{c|}{\textbf{HumanEval+}} & \multicolumn{3}{c}{\textbf{MBPP}} \\
 & \textbf{\passone} & \textbf{\passtwo} & \textbf{\passfive} & \textbf{\passone} & \textbf{\passtwo} & \textbf{\passfive} & \textbf{\passone} & \textbf{\passone} & \textbf{\passfive} \\
\hline
GPT-4o & 84.67 & 89.69 & 92.5 & 73.82 & 81.03 & 85.80 & 71.19 & 76.49 & 79.89 \\
\ourtooloo{} (GPT-4o-Based) & 94.96 & 96.10 & 97.64 & 81.95 & 86.4 & 90.9 & 75.36 & 75.94 & 76.20 \\
\ourtoolgt{} (GPT-4o-Based) & 97.59 & 99.15 & 99.03 & 85.50 & 88.42 & 91.58 & 85.30 & 86.56 & 88.13 \\
\hline
\oo{} & 90.96 & 92.59 & 93.95 & 93.60 & 95.63 & 96.00 & - & - & - \\
\ourtoolgt{} (o1-Based) & 100.0 & 100.0 & 100.0 & 95.26 & 99.95 & 100.0 & - & - & - \\
\hline
\end{tabular}
\end{adjustbox}

\end{table}

%% file: sections/rq2.tex
\subsection{RQ2: Users Efforts and User Study}
\label{sec:user_study}

Since \ourtool{} is designed as an interaction framework, it is essential to minimize the need for developer intervention. In this section, we focus mainly on the effort required from users. First, we show the interaction rounds in the simulated experiments to reflect user effort from a quantitative perspective. Additionally, we conduct a user study to evaluate \ourtool{} in interactive scenarios to evaluate the efforts from user experience.

\subsubsection{Experimental design of User Study}
In this section, we introduce the design of the user study.

\noindent \textbf{\emph{Settings.}} In our user study, there are three different settings. (1) \textit{Writing Code.} In the first setting, given the problem instruction in natural language, the users are asked to implement the function completely. (2) \textit{Fixing Code.} In the second setting, given the instruction, we will leverage the \llm{} to generate a code. Then the instruction and the code will be given to the users, and they will be asked to check the correctness of the code and fix it if any bugs exist. (3) \textit{Fixing Tests.} The third setting is \ourtool{}, and we will let the users use \ourtool{}. Each iteration \ourtool{} will identify a problematic test and ask the users to check it and fix it if any problem exist.

\noindent \textbf{\emph{Metrics.}} During the three settings, we will record the time that the users spend on each setting and ask the users to give a difficulty score representing the difficulty degree to complete the task in each setting (1 is easiest, and 5 is highest). Finally, we will compute \passone{} for the solutions obtained from each setting.

\noindent \textbf{\emph{Procedure.}} We randomly sample 20 examples from \humaneval{} and \mbpp{}. The implementation language is Python, and we select 6 PhD students who have three to five years of Python programming experiences. They do not learn about the problems in \humaneval{} and \mbpp{} before. We will have three different users complete each of the three settings for a given problem to prevent any user from becoming familiar with the problem after working on one setting.

\input{tables/iterations}
\input{tables/user_study}
\subsubsection{Experimental Results}
Firstly, we present the interaction rounds in the simulated experiments in Table~\ref{tab:iterations}. The average number of rounds across all benchmarks is 4.53, indicating that users need to be consulted only four times on average to achieve a 33\% performance improvement over the base model, which is a worthwhile efforts. 
\humaneval{}  and \humanevalplus{} require only 2.83 and 4.80 rounds, a relatively low number. In contrast, \mbpp{} involves more rounds, that is, 5.95. As discussed in RQ1, \mbpp{}'s instructions are somewhat ambiguous, making it more challenging for models to correctly refine the code, thus requiring additional tests for better understanding. 

Next, we present the results of the user study, summarized in Table~\ref{tab:user_study}. These results are averaged across all 20 problems. Among the three metrics evaluated, \ourtool{} (Fixing Tests) achieves the best performance.
For time spent and difficulty scores, users spend the least time on \ourtool{} and give it the lowest difficulty scores, which is consistent with our expectations. This can indicate the lightweight user efforts from the perspective of user experirence. For the \passone{}, we observed interesting results. The \passone{} of \ourtool{} and directly writing the code are the same and higher than fixing the code. This is likely because developers may overlook some bugs when given code to fix, but they are less likely to make the same mistakes when writing code from scratch or fixing test cases. Comparing \ourtool{} with writing code directly, the \passone{} rates are the same, but the time spent on \ourtool{} is much less, indicating that \ourtool{} is more efficient than writing code directly.
To confirm our observations, we further conducted two Wilcoxon signed-rank tests on time consumption between \ourtool{} and the other two methods (i.e., writing code and fixing code). The results of the tests show a p-value of 0.037 between \ourtool{} and writing code and a p-value of 0.033 between \ourtool{} and fixing code, indicating that \ourtool{} is statistically more time-efficient at a 95\% confidence level for users.


%% file: tables/iterations.tex
\begin{table}[t]
        \caption{The number of interaction rounds}
        \label{tab:iterations}
    \centering
\begin{tabular}{lcccc}
    \toprule
    \textbf{Benchmark}& \textbf{HumanEval} & \textbf{HumanEval+} & \textbf{MBPP} & \textbf{Average} \\
    \midrule
    \ourtool{} & 2.83 & 4.80 & 5.95 & 4.53 \\
    \bottomrule
\end{tabular}

    \end{table}


%% file: tables/user_study.tex
\begin{table}[ht]
        \caption{The results of user study}
        \label{tab:user_study}
        \centering
        \begin{tabular}{lccc}
            \toprule
            \textbf{Settings} & \textbf{Pass@1} & \textbf{Time} & \textbf{Difficulty} \\
            \midrule
            Writing Code  & \textbf{90} & 240.1 & 2.55 \\
            Fixing Code   & 80 & 156.4 & 2.10 \\
            Fixing Tests  & \textbf{90} & \textbf{120.9} & \textbf{1.65} \\
            \bottomrule
        \end{tabular}

    \end{table}

%% file: sections/rq4.tex
\subsection{RQ3: Time and Cost Overhead}
\label{sec:cost}

\input{tables/o1.tex}

\subsubsection{Experimental Design.}
In this RQ, we analyze the overhead of \ourtool{} in terms of time and cost. The primary expense for \ourtool{} arises from invoking the \openai{} API. The tool comprises three stages: generating code, correcting tests, and fixing code. Since the human efforts are already studied in the user study, we mainly focus on evaluating the time and cost of \ourtooloo{} in this RQ. We present results for \humaneval{} and \mbpp{} only since the problem instructions in \humanevalplus{} and \humaneval{} are identical. To provide intuitive comparisons, we also report the time and cost of using \oo{} solely without \ourtool{}. Given the substantial costs of \oo{}, when evaluating \oo{}, we sample 25 problems, generating 50 code candidates per problem using \oo. Results are summarized in Table~\ref{tab:cost}.

\subsubsection{Experimental Results.}
As shown in Table~\ref{tab:cost}, \ourtool{} achieves an overall runtime of 1 to 2 minutes, with a total cost of only \$0.30 to \$0.56. Among the three tasks, test correction demands most time and cost, as it relies on \oo{} and other two tasks use \gptthree{}. The \oo{} API is approximately 20 times more expensive than the \gptthree{} API. However, because \ourtool{} requires only about four rounds of feedback, the overall cost of invoking \oo{} remains manageable. In comparison, using solely \oo{} results in an average runtime of 25.44 minutes and a cost of \$5.87 per problem, exceeding \ourtool{}'s cost by over tenfold. Moreover, \ourtool{} even outperforms \oo{}. Thus, \ourtool{} efficiently achieves superior performance with minimal time and budget compared to the most advanced model alone.

%% file: tables/o1.tex

\begin{table}[t]
\caption{Time and cost analysis of \ourtool{} and OpenAI o1.}
\label{tab:cost}
\centering
\begin{adjustbox}{width=\columnwidth}
\begin{tabular}{ll|ccc|cccc|cccc}
\toprule
\multirow{2}{*}{\textbf{Models}} & \multirow{2}{*}{\textbf{Datasets}}& \multicolumn{3}{c|}{\textbf{Performance}} & \multicolumn{4}{c|}{\textbf{Avg Time (min)}} & \multicolumn{4}{c}{\textbf{Avg Cost (\$)}} \\
& & \textbf{Pass@1} & \textbf{Pass@2} & \textbf{Pass@5} & \textbf{Overall} & \textbf{Gen} & \textbf{Correct} & \textbf{Fix} & \textbf{Overall} & \textbf{Gen} & \textbf{Correct} & \textbf{Fix} \\
\midrule
\ourtooloo{} & MBPP & 73.68 & 74.66 & 75.76 & 1.08 & 0.21 & 0.48 & 0.39 & 0.30 & 0.03 & 0.21 & 0.06 \\
(\gptthree-based)& HumanEval & 92.00 & 92.00 & 95.98 & 2.26 & 0.28 & 1.10 & 0.88 & 0.56 & 0.04 & 0.40 & 0.12 \\
\midrule
\oo & HumanEval & 90.96 & 92.59 & 93.9 & 25.44 & 25.44 & - & - & 5.87 & 5.87 & - & - \\
\bottomrule
\end{tabular}
\end{adjustbox}

\end{table}

%% file: sections/discussion.tex
\noindent \textbf{\emph{Integrating User Feedback}}
We introduce a lightweight interaction framework to incorporate user feedback for test correction. Human involvement is essential for two main reasons. First, in many cases, only the developer can verify the correctness of specific test results. Since both tests and code are generated by LLMs, they may each contain inaccuracies, making it challenging to determine correct test outputs from these sources alone. In real-world development, developers must write tests to ensure the correctness of the codes. Second, maintaining the developer's involvement fosters a sense of code ownership and ensures they maintain a clear understanding of the code generation process. In software development, the process itself often carries more significance than the final result. This engagement allows developers to gain a deeper understanding of code details, enhancing their ability to address future bugs swiftly. However, it is important to limit developer involvement. We ask users only to validate test results, not the code itself. \ourtool{} can achieve a 33\% improvement in performance with only four rounds of interaction. Additionally, we propose an automated variant, \ourtooloo{}, which also outperforms the standard \sotasimple{} approach.

\noindent \textbf{\emph{Threats to Validity}}
\textit{Threats to Internal Validity} mainly lie in the randomness introduced by \llm{}s. To address this issue, we use identical code and tests generated by \llm{}s for different techniques to maintain consistency. Additionally, for fair comparison, we employ the same ChatGPT API as the \sota{} technique \mpsc{}~\cite{huang2023enhancing}. To ensure accuracy, the first two authors meticulously review the code to prevent bugs.
\textit{Threats to external validity} mainly lie in the benchmark used. To prevent data leakage issues with \llm{}s, we use hand-written datasets that are not part of the \llm{} training data. These datasets are consistently used across all code generation techniques. Another external validity threat arises from the user study. Each problem requires developers to perform three different tasks, and to minimize bias, we assign different developers to each task within a problem. This introduces the potential threat that the three developers may have varying levels of expertise. To mitigate this, we select users with comparable development experience.



\noindent \textbf{\emph{Limitations}}  The first limitation of \ourtool{} is that its performance depends on the quality of the corrected tests. If tests are incorrectly corrected, the model may be misled—a limitation shared by all human-involving techniques. However, correcting tests is generally less error-prone than directly correcting code. Additionally, in real-world development, users must ensure the correctness of tests; otherwise, the quality of the code may be compromised. While formal verification could be used to ensure test correctness, it is often too time-consuming for practical use.

%% file: sections/relatedwork.tex
In this section, we present the related work relevant to \ourtool{}. We cover three types of related work: \llm{} for code generation, \self{}, and other post-hoc techniques.

\noindent \textbf{\emph{\llm{} for Code Generation}}
Code generation techniques automatically produce code snippets based on natural language descriptions, which has high practical value and has been studied for decades. Early methods relied on templates to generate code~\cite{zettlemoyer2007online, zettlemoyer2012learning, kushman2013using}. With the appearance of neural networks, many learning-based techniques~\cite{yin2017syntactic, rabinovich2017abstract, sun2020treegen} emerged, using natural language requirements as input and generating code as output.
Recently, \llm{}s have shown strong performance in various tasks, including code generation. General-purpose \llm{}s, such as \chatgpt{}\cite{gpt4o} and Claude\cite{claude}, are trained on diverse types of textual corpora, achieving high performance across NLP tasks, including code generation. In addition, code-specific \llm{}s, trained on extensive public code datasets, such as DeepSeek-Coder~\cite{guo2024deepseek}, WizardCoder~\cite{luo2023wizardcoder}, and StarCoder~\cite{li2023starcoder}, have been developed. All these techniques utilize a similar network architecture based on the Transformer model~\cite{vaswani2017attention}, with only minor modifications, such as adjustments in layer count and positional embeddings. These techniques commonly adopt a two-stage training process: (1) next token prediction with ``fill-in-the-middle'' (FIM) and (2) instruction tuning. FIM enables models to predict masked snippets using surrounding code, enhancing their understanding of code context and generation ability. Instruction tuning further improves the model's capability to follow human instructions, essential for generating code based on natural language commands.

\noindent \textbf{\emph{\self{}}}
Despite achieving promising performance, \llm{} outputs remain unreliable, particularly for complex tasks where we cannot guarantee output accuracy. This requires users to verify \llm{} outputs, producing additional human efforts. To improve output reliability, researchers have proposed leveraging \self{} to post-process results. By sampling multiple outputs from \llm{}s and using majority voting, the approach selects the most consistent response as the final answer. This \self{} method is theoretically based on the notion that the tasks may allow multiple valid paths to the correct answer~\cite{stanovich2000advancing}. Rooted in the principle of \diversity{}, \self{} assumes that when diverse reasoning methods converge on a single answer, it is more likely accurate. 
For instance, \citet{wang2022self} samples multiple ``chain-of-thought'' paths, choosing the most frequent answer, while \citet{sun2022recitation} generates varying outputs by first reciting different relevant knowledge. These methods judge consistency directly from output agreement.
However, code generation is an open-ended problem, so \self{} techniques in code generation domain rely on different \indicator{}s to assess consistency. \codet{}\cite{chen2022codet} generates both codes and multiple tests, selecting the code that passes the most tests (inter-consistency) and shares the most \functionequal{} \samecode{}s (intra-consistency). Both types of consistency depend heavily on test quality, yet these tests, generated by \llm{}s, may contain errors. \mpsc{}\cite{huang2023enhancing} considers \spec{} alongside tests, though this approach also faces limitations. As shown in their paper~\cite{huang2023enhancing}, \spec{} achieves only around 50\% accuracy, and has small impact on results after removing \spec{}. To address these issues, \ourtool{} enhances the quality of tests and codes iteratively through a co-evolution process. Other consistency types involve cross-model consistency, where different \llm{}s debate when their answers are inconsistent~\cite{xiong2023examining}. ALGO~\cite{zhang2023algo} introduces an additional brute-force implementation, using its output as an oracle to verify the correctness of other generated outputs. However, the accuracy of this brute-force algorithm itself is not guaranteed.

\noindent \textbf{\emph{Other Post-Process Techniques}}
Besides \self{}, there are also other post-process techniques improving the quality of \llm{}s' outputs. Self-refine~\cite{madaan2024self} prompts the \llm{} to assess its output, provide feedback, and refine based on that feedback. However, these methods lack explicit guidance for modification, requiring the \llm{} to self-reflect, which can be challenging. In contrast, we provide \llm{}s with failed test cases as direct feedback. Reflexion~\cite{shinn2023reflexion} uses environmental feedback, such as error messages during test execution, to refine outputs, though the authors acknowledge that performance relies on test quality. Other techniques~\cite{ni2023lever,zhang2023coder} introduce a separate verifier or reviewer to score and re-rank outputs, yet these approaches also lack a specific direction for evaluation. In our work, tests offer the most precise guidance for evaluation and refinement.

%% file: sections/conclusion.tex

\llm{}s have shown promising performance in code generation; however, they struggle to produce flawless code in a single attempt. Researchers leverage \self{} to enhance code quality. Nevertheless, current methods overlook a crucial aspect of using \self{}: the \indicator{}s should be of good quality. Without this, the achieved consistency remains unreliable. 
In this work, we introduce Consistency-Augmented Iterative Interaction
Framework to Enhance the Reliability of Code Generation (\ourtool{}), an approach designed to enhance code generator performance through two key components: (1) lightweight user effort for validating the correctness of selected tests, and (2) a dynamic strategy for ranking, localizing, and correcting multiple tests and codes.
Our framework enables a lightweight, interactive process that incorporates user feedback to address identified tests and guide the iterative improvement process. Notably, the iteration rounds average only four with the support of consistency, requiring minimal human effort to achieve a performance improvement of approximately 30\%.
Each iteration follows a co-evolution process involving codes and tests. This process iteratively refines code and test quality, making consistency voting from codes to tests and vice versa increasingly reliable.